\documentclass[12pt]{article}

\usepackage{cite}
\usepackage{amsmath,amssymb,amsfonts}
\usepackage{algorithmic}
\usepackage{graphicx}
\usepackage{textcomp}
\usepackage{xcolor}
\usepackage{url}
\usepackage{booktabs}
\usepackage{multirow}
\usepackage[margin=1in]{geometry}

\usepackage{cite}
\usepackage{amsmath,amssymb,amsfonts}
\usepackage{algorithmic}
\usepackage{graphicx} 
\usepackage{textcomp}
\usepackage{xcolor}
\usepackage{url}
\usepackage{booktabs}
\usepackage{multirow}

\begin{document}



\title{Validating Behavioral Proxies for Disease Risk Monitoring via Large-Scale E-commerce Data}

\author{%
Naomi Sasaya\\
LY Corporation\\
Tokyo, Japan
\and
Shigefumi Kishida\\
Anicom Insurance, Inc.\\
Tokyo, Japan
\and
Ryo Kikuchi\\
Anicom Insurance, Inc.\\
Tokyo, Japan
\and
Akira Tajima\\
LY Corporation\\
Tokyo, Japan
}

\date{}

\maketitle


\begingroup
\hyphenpenalty=10000
\exhyphenpenalty=10000

\begin{abstract}
\textbf{Background:}
Digital traces of daily activities, such as online search queries and purchase histories, are gaining attention as alternative data sources for public health surveillance. However, their practical application remains limited because such behavioral signals often do not directly correlate with clinically confirmed disease outcomes.

\textbf{Objective:}
This study aims to assess the epidemiological validity of a scalable behavioral proxy for disease risk monitoring derived from large-scale e-commerce (EC) purchase data by comparing it with independent clinical insurance records.

\textbf{Methods:}
We defined a behavioral proxy for disease onset using EC purchase logs. We then evaluated the utility of this proxy using two complementary sources of insurance-derived ground truth data, together with EC purchase data ($N = 55{,}645$ users). Specifically, ingredient-level risk patterns and seasonal disease dynamics were assessed.

\textbf{Results:}
Proxy-based risk estimates showed strong agreement with clinical data in feline lower urinary tract disease (FLUTD), with a correlation of $r = 0.74$ for ingredient-level risk patterns and $r = 0.82$ for seasonal variation. Both data sources showed consistent seasonal trends, with increased disease risk observed during winter months. In addition, analysis based solely on EC purchase data showed that higher wet food consumption was associated with lower estimated disease risk, consistent with established veterinary findings.

\textbf{Conclusions:}
These results suggest that validated behavioral signals derived from EC purchase data may provide a complementary data source to traditional public health surveillance systems in a cost-effective and scalable manner. The proposed framework may be applicable beyond veterinary health, including the surveillance of lifestyle-related and chronic diseases in human populations.

\end{abstract}
\endgroup



\section{Introduction}
In the fields of healthcare and epidemiology, ``digital epidemiology''---the use of web-based behavioral logs, often referred to as lifelogs, to estimate health status and disease risk---has gained significant attention. Predicting influenza trends using search queries or social media posts is a well-known example of this approach. More recently, purchase histories from e-commerce (EC) platforms have also been recognized as valuable data sources that reflect individuals’ long-term lifestyle habits.

However, utilizing EC data and other web-based logs for medical risk analysis presents a fundamental challenge: these datasets typically lack ground truth labels based on clinical diagnoses. 
It is difficult to reliably distinguish, based on log data alone, whether a specific product purchase reflects preventive action, treatment after disease onset, or simply a change in consumer preference. Consequently, raw web data cannot be readily used as reliable inputs for epidemiological analyses without additional validation.

Traditionally, research investigating the relationship between diet and disease (dietary factors) has relied on clinical trials or questionnaire-based surveys. However, these approaches are costly, and it is challenging to follow tens of thousands of subjects over long periods. In contrast, EC data can be collected at large scale and low cost, but are limited by the lack of medical validation (ground truth).

To address this challenge, we propose an approach to assess the epidemiological reliability of EC purchase logs by cross-referencing EC purchase data with two complementary sources of insurance-derived data. Specifically, we define the behavioral proxy of disease onset as switching from regular diets to therapeutic diets on EC platforms. 
We then quantitatively evaluate the epidemiological validity of this proxy using two complementary sources of insurance-derived ground truth data. Specifically, ingredient-level risk patterns are assessed using questionnaire-linked insurance data for feline cystitis, while seasonal disease dynamics are evaluated based on aggregated insurance claim data for feline lower urinary tract disease (FLUTD). In addition, as a case study, we apply the validated proxy to EC purchase data alone to analyze the relationship between diet and disease risk in FLUTD.

The contributions of this study are twofold:
\begin{enumerate}
    \item We demonstrate the epidemiological utility of EC data by showing a statistically significant association between purchase-based behavioral signals and clinical outcomes.
    \item We show that established veterinary knowledge, such as the protective effect of wet food consumption, can be reproduced using web-based behavioral data once the proxy has been validated.
\end{enumerate}


\section{Related Work}
\subsection{Veterinary Background: FLUTD and Diet}
FLUTD is a common disease in cats, and diet plays a major role in its onset. Buffington et al. \cite{buffington1997} reported that cats developing idiopathic cystitis were significantly more likely to be fed only dry food compared to healthy cats. Case-control studies by Jones et al. \cite{jones1997} and Piyarungsri et al. \cite{piyarungsri2020} also indicated that dry-food-dominant diets are risk factors for FLUTD. While clinically important, these studies were often limited to sample sizes of tens to hundreds, creating a need for validation using large-scale data.

\subsection{Health Risk Analysis Using Web Data}
Estimating health status and disease risk using web-based behavioral data, such as search logs and purchase histories, has been widely studied in the informatics and data mining communities. 
A representative line of work leverages search query frequencies to monitor and predict infectious disease outbreaks, including influenza-like illness trends \cite{ginsberg2009}. 
Other studies have shown that large-scale purchase data, such as pharmaceutical transaction logs, can be used to detect adverse drug reactions and other health-related signals \cite{white2013}.

Building on prior exploratory work using EC data to analyze associations between pet food ingredients and allergic diseases in dogs \cite{sasaya2020}, this study extends that line of research by validating a behavioral proxy against independent clinical records. By cross-referencing behavioral signals derived from purchase logs with insurance claim data used as ground truth, we provide an objective assessment of the proxy’s epidemiological reliability, which has been largely missing in prior web-based health studies.


\section{Datasets}
We used two main categories of datasets: EC purchase data and insurance-derived data, both of which were statistically processed to prevent the identification of individuals or specific animals, in accordance with the security guidelines and ethical regulations of the respective companies.

\subsection{EC Purchase Data (Proxy)}
We used purchase logs from Yahoo! Shopping, provided by Yahoo Japan Corporation (now LY Corporation), covering a three-year period from January 2018 to December 2020. Although these data precede the present, the physiological mechanisms of cats and the causal relationships between dietary ingredients and disease risk are biological factors that are unlikely to be affected by short-term temporal changes. Therefore, we consider the verification results from this period to be applicable to current settings.

\subsubsection{Target Product Definition}
Target products, which serve as signals of disease onset, were defined as products in the therapeutic or semi-therapeutic diet category that contained specific keywords (e.g., FLUTD, Lower Urinary Tract, pH Control) in their product names (Table~\ref{tab:keywords}).

\begin{table}[htbp]
 \caption{Keywords for Target Product Extraction}
 \label{tab:keywords}
 \begin{center}
  \begin{tabular}{ll}
   \toprule
   Category & Keywords (examples) \\
   \midrule
   Disease Name & FLUTD, Lower Urinary Tract, Urinary Disease \\
    & Struvite, Stone \\
   Function/Ingredient & Urinary, pH Control, pH Care \\
    & pH Balance, Mineral Control \\
   \bottomrule
  \end{tabular}
 \end{center}
\end{table}

\subsubsection{Analysis Subjects}
Users who purchased general cat food products (1,271 products) distributed in Japan were classified into two groups:
\begin{itemize}
    \item \textbf{Cases (Estimated Onset Group)}: Users who switched to purchasing target products for the first time after previously purchasing general food ($N = 4{,}328$). The analysis focused on the one-year period prior to the first target purchase.
    \item \textbf{Controls (Estimated Healthy Group)}: Users who purchased only general food and never purchased target products ($N = 51{,}317$). The analysis focused on the most recent one-year period.
\end{itemize}

\subsection{Insurance Claim Data (Ground Truth)}
We used two distinct insurance-derived datasets provided by Anicom Insurance, Inc., each serving a different analytical purpose in this study.

\subsubsection{Questionnaire-linked case--control dataset (Ingredient risk analysis)}
This dataset was used for the ingredient-level risk analysis. It is based on questionnaire responses on dietary intake from cats that filed insurance claims for feline cystitis (Case Group, $N = 296$) and from cats that had never filed insurance claims (Control Group, $N = 9{,}158$).

\subsubsection{Aggregated insurance claims dataset (Seasonality analysis)}
For the seasonality analysis, we used monthly insurance claim counts for FLUTD over a three-year period. Because insured population denominators were unavailable, the data were analyzed as claim counts rather than population-normalized incidence rates.

\section{Methodology}
\begin{figure}[htbp]
 \centering
 \includegraphics[width=0.95\linewidth]{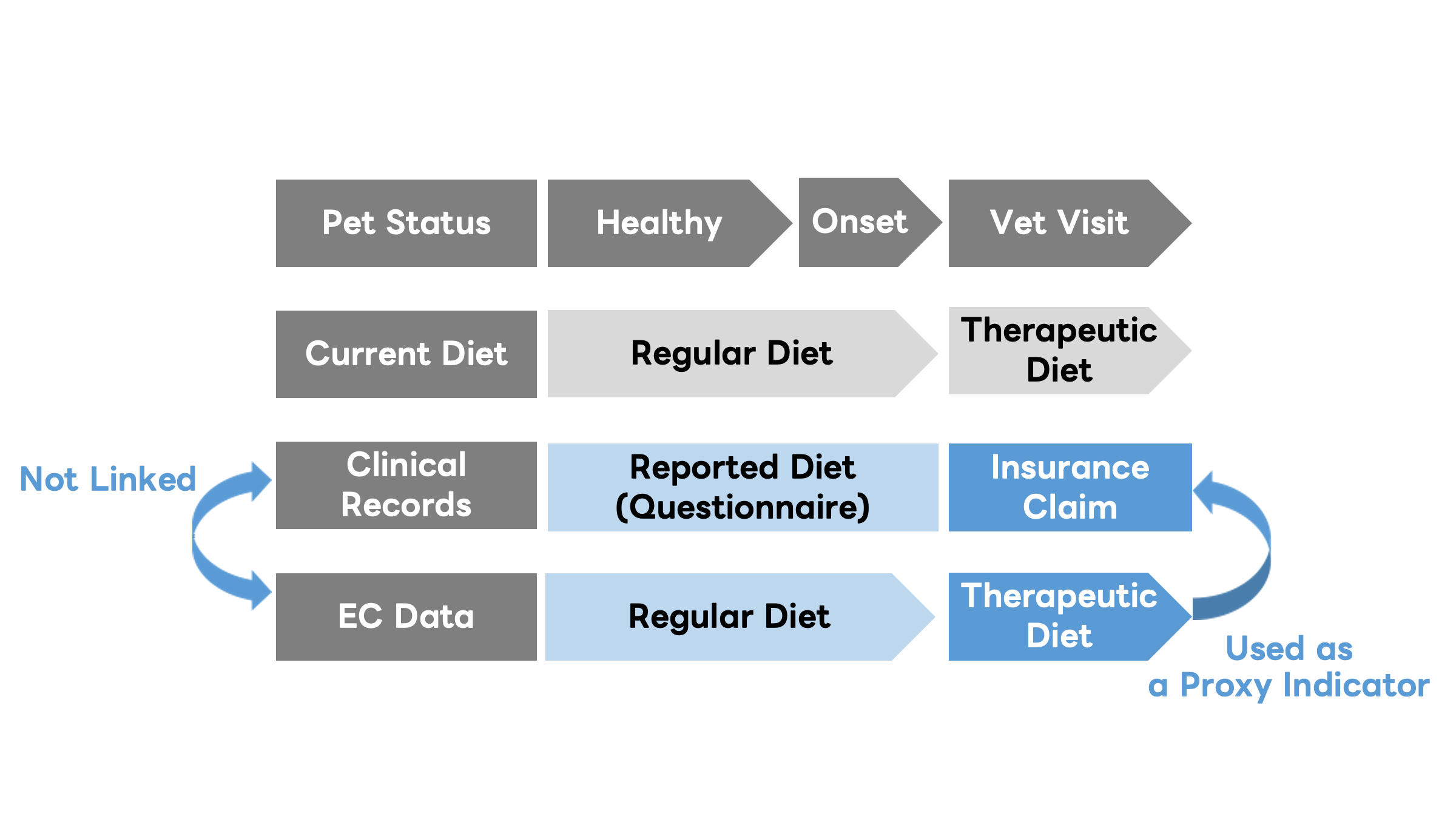}
 \caption{Correspondence between clinical records and EC purchase data. The two data sources are not directly linked at the individual level.}
 \label{fig:concept}
\end{figure}

The deterioration of a pet’s health in clinical settings is reflected in veterinary visits (insurance claims) and subsequently in dietary management at home (therapeutic diet purchases). To capture this behavioral modification, we define the following indicators (Fig.~\ref{fig:concept}).

\subsection{Proxy Indicator: Switch Rate}
In the EC data, the disease risk for a user group with a certain attribute (e.g., consuming specific ingredients) is defined as the transition rate (Switch Rate) from general food products to therapeutic diets.
\begin{equation}
    \text{Switch Rate} = \frac{\text{Cases}}{\text{Cases} + \text{Controls}}
\end{equation}

In this study, the term "general food" refers to EC product categories, whereas "regular diet" is used to describe habitual feeding behavior prior to switching to therapeutic diets.

\subsection{Ground Truth Indicator: Claim Rate}
The risk indicator derived from insurance data, serving as the comparison standard, is defined as follows:
\begin{equation}
    \text{Claim Rate} = \frac{\text{Case Group}}{\text{Case Group} + \text{Control Group}}
\end{equation}

\begin{figure}[htbp]
 \centering
 \includegraphics[width=0.90\linewidth]{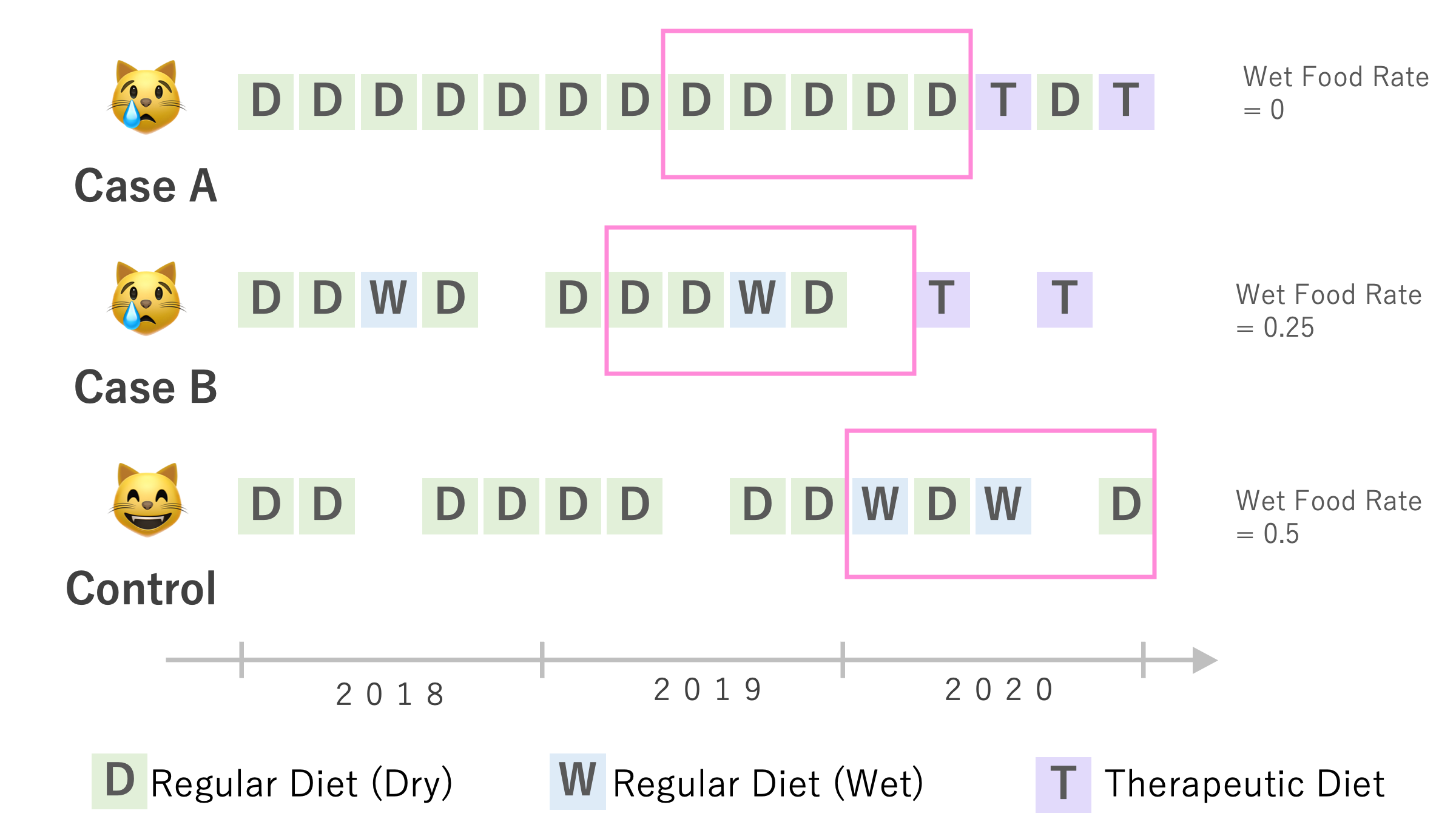}
 \caption{Definition of Case Group and Control Group extracted from time-series insurance records.}
 \label{fig:definition}
\end{figure}

The procedure for extracting the Case Group and Control Group from time-series insurance records is illustrated in Fig.~\ref{fig:definition}.

\section{Evaluation: Validity of the Proxy}
To assess whether the proposed proxy (Switch Rate) reflects the true risk (Claim Rate), we evaluated it from two perspectives: ingredient correlation and seasonal variation.

\subsection{Correlation Analysis Based on Ingredients}
\begin{figure}[htbp]
 \centering
 \includegraphics[width=1.0\linewidth]{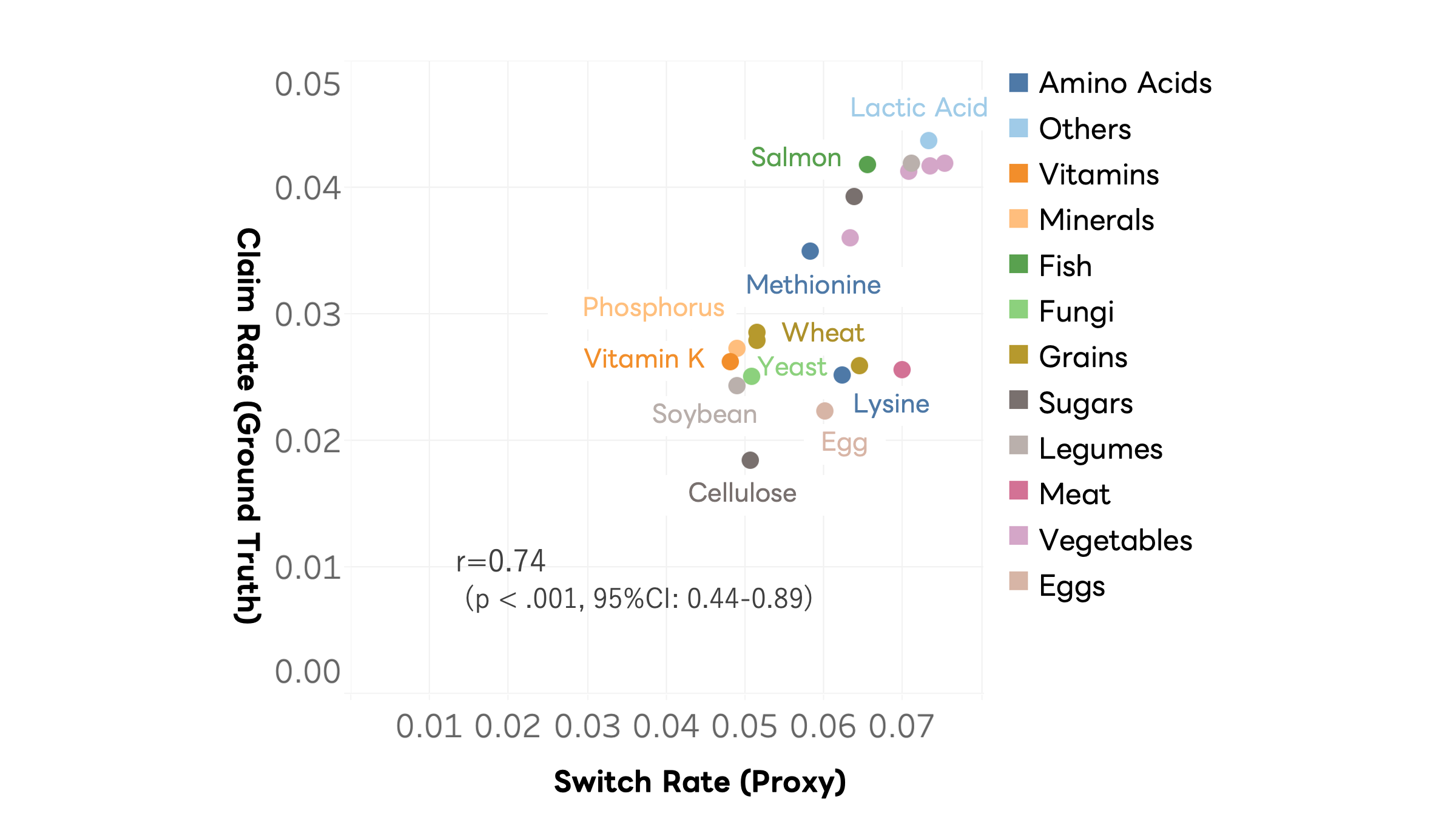}
 \caption{Correlation between Claim Rate (Ground Truth) and Switch Rate (Proxy) across ingredient categories ($r = 0.74$).}
 \label{fig:correlation}
\end{figure}

We calculated risk indicators in both datasets for each individual ingredient (411 types) contained in the foods.
For major ingredient categories where a significant difference ($p<0.05$) was found by the Chi-squared test, we created a scatter plot (Fig.~\ref{fig:correlation}). The correlation coefficient was $r=0.74$ ($p < 0.001$), confirming a strong positive correlation.
This indicates that groups consuming dietary ingredients considered high-risk in clinical practice also transition to therapeutic diets with high probability in the EC data, suggesting that the behavioral proxy captures risk trends at the ingredient level.

\subsection{Verification of Seasonal Variation}
FLUTD is known to be a seasonal disease with increased onset in winter. Therefore, we performed STL decomposition (Seasonal-Trend decomposition using Loess) on the monthly time-series data of aggregated insurance claim counts for FLUTD and the number of first-time therapeutic diet purchasers on EC platforms to compare their seasonal components.

\begin{figure}[htbp]
 \centering
 \includegraphics[width=0.95\linewidth]{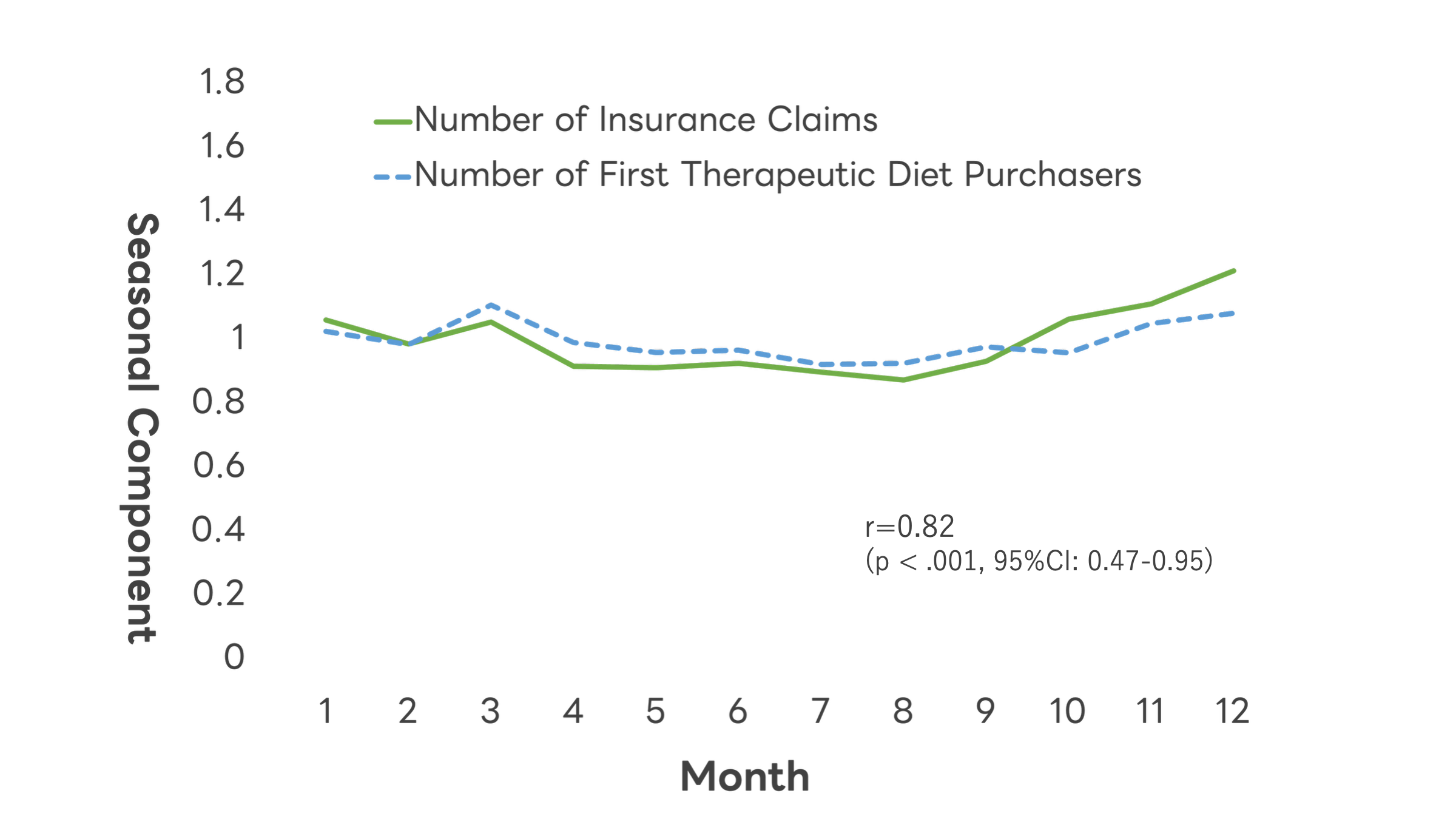}
\caption{Comparison of seasonal components of FLUTD insurance claims and first-time therapeutic diet purchasers on EC platforms ($r = 0.82$).}
 \label{fig:seasonality}
\end{figure}

The results showed a correlation coefficient of the seasonal components of $r=0.82$ ($p < 0.001$; Fig.~\ref{fig:seasonality}). 
Both datasets consistently peaked in winter (December to January), indicating that EC data capture real-world disease trends.
Importantly, no systematic temporal lag was observed between the two signals.

\section{Case Study: Wet Food and Disease Risk}
Having confirmed the validity of the proxy indicator, we present an application of the proxy using only EC data to analyze disease risk at scale. We investigated the relationship between dietary moisture content (wet food rate) and FLUTD risk.

\subsection{Results}
\begin{figure}[htbp]
 \centering
 \includegraphics[width=0.95\linewidth]{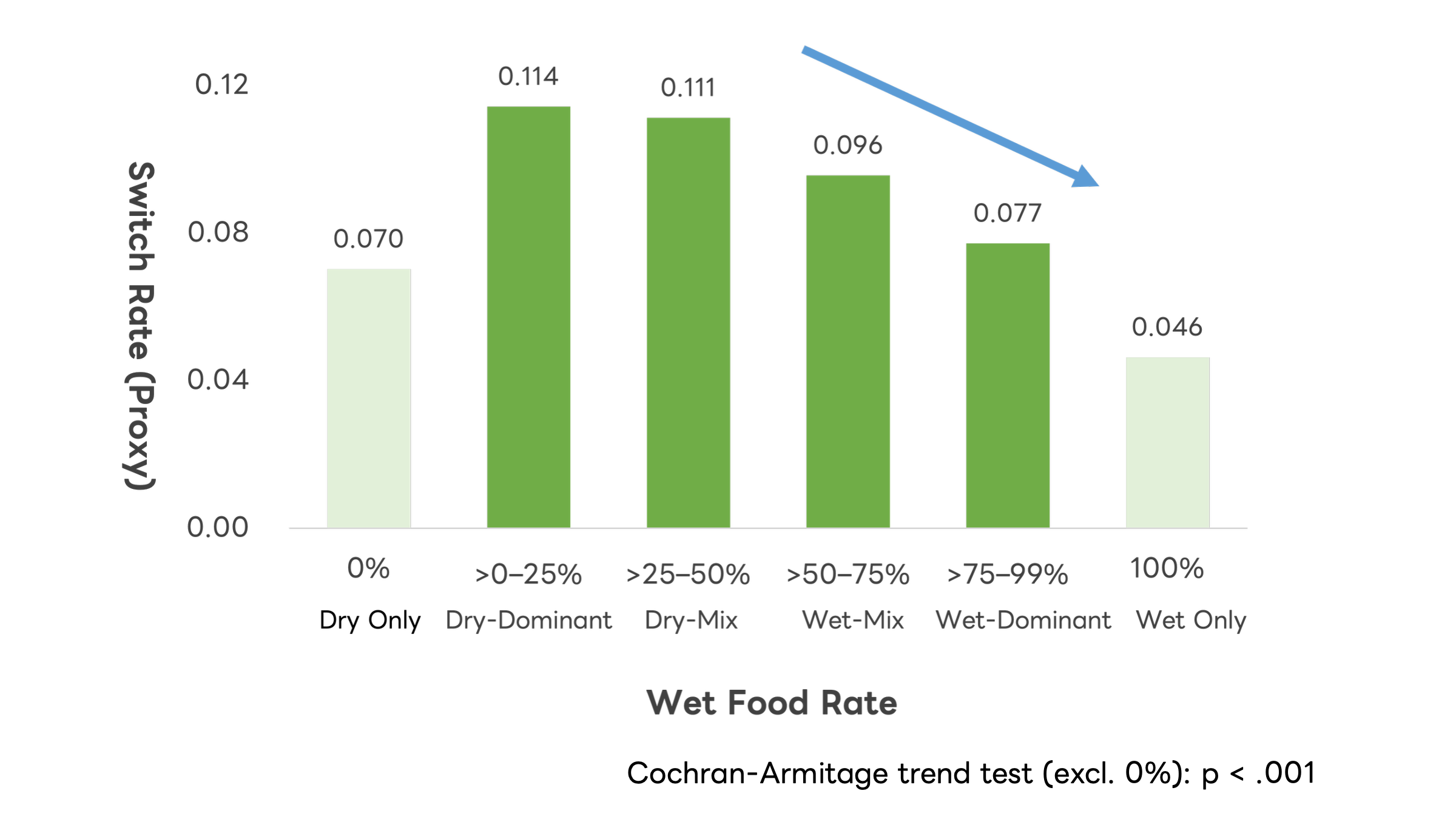}
 \caption{Relationship between wet food rate and disease risk (Switch Rate).}
 \label{fig:case_study}
\end{figure}

For each user in the EC data, we calculated the wet food rate as the proportion of wet food among purchased regular diets, and computed the Switch Rate for each group. As shown in Fig.~\ref{fig:case_study}, excluding users who exclusively fed dry food (0\% wet food), the Switch Rate exhibited a significant decreasing trend as the wet food purchase ratio increased (Cochran--Armitage trend test, $p < 0.001$), indicating a dose--response relationship. Users in the dry-dominant group ($\leq 25\%$) showed an estimated onset risk approximately 1.48 times higher than users in the wet-only group (100\%).

\section{Discussion}
\subsection{Reproducibility of Medical Knowledge via Web Data}
The analysis revealed a trend in which higher wet food rates were associated with lower disease risk, consistent with findings in veterinary medicine. The ability to reproduce established medical knowledge using only web-based data suggests that the proposed method may be applicable to the discovery of previously unknown disease risk factors.

\subsection{Limitations and Bias}
The 0\% wet food group should not be interpreted as a low-risk group, but rather as a reflection of behavioral and observational biases. As shown in Fig.~\ref{fig:case_study}, the Switch Rate for the exclusively dry-food group (0\% wet food) was lower than that of the mixed-feeding group ($\leq 25\%$ wet food), which requires careful interpretation. One plausible explanation is survivorship bias, whereby cats that continue to consume only dry food may represent individuals that did not experience adverse health outcomes under that diet. In addition, latent disease processes that remain unrecognized by owners may not prompt changes in purchasing behavior, leading to an underrepresentation of disease risk in purchase-based proxies. Differences in EC usage intensity or information-seeking behavior between owners who rely exclusively on dry food and those who adopt mixed feeding practices may also contribute to the observed patterns.

Finally, the seasonality analysis is based on aggregated insurance claim data and EC purchase behavior rather than population-normalized incidence rates. Therefore, the observed seasonal patterns should be interpreted with caution, taking into account non-clinical factors and limitations inherent to aggregated claim data.

\subsection{Implications for Public Health Surveillance}
Although this study focuses on veterinary health as a case example, its primary contribution lies in the validation framework for behavioral proxies rather than in the specific disease domain. Behavioral data derived from digital platforms have been widely explored for monitoring human health conditions, including infectious diseases and medication-related events \cite{ginsberg2009,white2013}. 
Our findings suggest that, once a behavioral signal is quantitatively validated against independent clinical data, such signals can provide useful insights into population-level disease risk.
This framework may be particularly relevant for public health surveillance contexts in which timely access to clinical diagnoses is limited, delayed, or costly. Accordingly, the proposed approach has the potential to complement traditional surveillance systems for human health, especially for lifestyle-related and chronic conditions in which behavioral changes often precede formal medical diagnoses.

\section{Conclusion}
In this study, we proposed a behavioral proxy based on switching behavior to therapeutic diets in large-scale EC data and examined its epidemiological validity through comparison with independent insurance claim records. The results suggest that this proxy provides useful insights into population-level disease risk patterns that are consistent with clinical data, both in terms of ingredient-related risk and seasonal dynamics. While the proposed approach does not replace traditional clinical surveillance, it may serve as a complementary data source for large-scale disease risk monitoring, particularly in settings where access to timely clinical diagnoses is limited. Future work will explore the applicability of this framework to other diseases and domains, including potential extensions to human health surveillance.

\section*{Conflicts of Interest}
The authors are employees of LY Corporation and Anicom Insurance, Inc., which provided the data used in this study. The analysis was conducted for research purposes, and the authors declare that these affiliations did not influence the study design, analysis, or interpretation of the results.


\section*{Acknowledgments}
During the preparation of this manuscript, generative AI tools were used solely for language editing and proofreading. The study design, data analysis, interpretation of the results, and conclusions were conducted entirely by the authors.

\bibliographystyle{unsrt}
\bibliography{references}

\end{document}